\documentstyle[seceq,twocolumn,epsf]{jpsj}
\def\v#1{\mib #1}
\def\deltac{\delta_{\mbox{c}}}
\def\Dc{D_{\mbox{c}}}
\def\simleq{\mbox{\raisebox{-1.0ex}{$\stackrel{<}{\sim}$}}}
\def\simgeq{\mbox{\raisebox{-1.0ex}{$\stackrel{>}{\sim}$}}}

\def\runtitle{Critical Properties of Spin-1 with Bond Alternation and Uniaxial Single-Ion-Type Anisotropy}
\def\runauthor{Wei {\sc Chen}, Kazuo {\sc Hida} and Bryan C. {\sc Sanctuary}}

\title
{
Critical Properties of Spin-1 Antiferromagnetic Heisenberg Chains with Bond Alternation and Uniaxial Single-Ion-Type Anisotropy
}

\author
{
Wei {\sc Chen}\footnote{E-mail: chenwei@riron.ged.saitama-u.ac.jp}, Kazuo {\sc Hida} and Bryan C. {\sc Sanctuary}$^1$
}

\inst
{
Department of Physics, Saitama University, Urawa, Saitama, Japan 338-8570\\
$^1$ Department of Chemistry, McGill University, Montreal, PQ, Canada H3A 2K6 
}

\recdate
{
\today
}

\abst
{
One dimensional $S=1$ antiferromagnetic Heisenberg chains with bond alternation and uniaxial single-ion-type anisotropy are studied by numerical exact diagonalization of finite size systems. We calculate the ground state phase diagram using the twisted boundary condition method for $D\geq 0$. The ground states of this model contain the Haldane state, large-$D$ state and the dimer state. Using conformal field theory and the level spectroscopy method, we calculate the $D$ and $\delta$-dependence of the Luttinger liquid paraments $K$ and the critical exponent $\nu$ of the energy gap.
}

\kword
{
Heisenberg chain, exact diagonalization, conformal field theory, dimer phase, Haldane phase, large-$D$ phase
}

\begin{document}
\sloppy
\maketitle

\section{Introduction}
One dimensional antiferromagnetic Heisenberg chains have been the subject of recent investigations by numerous groups. The various phenomena of current interest are shown to be due to quantum effects. A uniform antiferromagnetic Heisenberg chain is known to have a gapless ground state for half-integer-spin. Especially, the exact solution is available for $S=1/2$ chain\cite{bh}. In contrast, for the integer-spin\cite{fd}, there is a gap between the first excited state and ground state.

One dimensional $S=1$ antiferromagnetic Heisenberg chains with bond alternation undergo a transition from the Haldane state to the dimer state as the strength of bond alternation increases. This transition has been studied by many methods\cite{ia,yt,kn}. On the other hand, the phase boundaries and critical properties of $S=1$ Heisenberg chains with uniaxial single-ion anisotropy\cite{gs} were determined by Glaus and Schneider using a finite size scaling method. The physical picture of both states was clarified by Nijs and Rommelse in terms of string order\cite{md} and by Tasaki using a stochastic geometric representation \cite{ht}. Using an exact diagonalization method and a phenomenological renormalization group approach, Tonegawa {\it et al.} calculated the phase diagram of the $S=1$ antiferromagnetic Heisenberg chain with bond alternation and uniaxial single-ion-type anisotropy. They found the Haldane phase, N\'{e}el phase, dimer phase and large-$D$ phase in the ground state. No phase transition was found, however, between the dimer phase and large-$D$ phase.\cite{tt} In this paper, we calculate the ground state phase diagram of a one demensional $S=1$ antiferromagnetic Heisenberg chain with bond alternation and uniaxial single-ion-type anisotropy using the twisted boundary condition method\cite{ak, kn}. This improves the accuracy of the transition points. Using conformal field theory and the level spectroscopy method, we obtain the Luttinger liquid parameter $K$ and the critical exponent $\nu$ of the energy gap.

This paper is organized as follows. In the next section, the model Hamiltonian is defined and the numerical results are presented. From the exact diagonalization data with the twisted boundary condition, the ground state phase diagram is obtained. The Luttinger liquid parameter $K$ is calculated by conformal field theory and the level spectroscopy method at the transition points. The final section is devoted to a summary and discussion.

\section{Numerical Results}
\subsection{Model Hamiltonian}
The Hamiltonian is given by
\begin{equation}
\label{de}
{\cal H} = \sum_{l=1}^{N} \{1-(-1)^{l} \delta \} \v{S}_{l} \v{S}_{l+1}+D\sum_{l=1}^{N}(S_{l}^{z})^{2},
\end{equation}
where $\v {S_{l}}$ is a spin-1 operator. The parameter $\delta(-1 \le \delta \le 1)$ and $D (D\ \geq 0)$ represent the bond alternation and uniaxial single-ion anisotropy, respectively. We take $\delta \geq 0$ without loss of generality. The ground state of this model can be regarded as a dimer state and large-$D$ state for large $\delta$ and $D$. For $D=0$, the ground state is the Haldane state or the dimer state depending as to whether $\delta<\deltac$ or $\delta>\deltac$ ($\deltac \cong 0.260 \pm 0.001$), respectively. For $\delta=0$, the ground state is the Haldane state or a large-$D$ state according to $D<\Dc$ or $D>\Dc$ ($\Dc \cong 1.001 \pm 0.001$), respectively. According to ref. \citen{tt}, there is no transition point between the dimer state and the large-$D$ state.

\subsection{Phase diagram for $D \geq 0$}
In order to determine the phase boundary  with high accuracy, we use the twisted boundary method of Kitazawa and Nomura \cite{ak,kn}. The Hamiltonian is numerically diagonalized to calculate the two low lying energy levels with the twisted boundary condition ($S^{x}_{N+1}=-S^{x}_1,$ $ S^{y}_{N+1}=-S^{y}_1,$ $ S^{z}_{N+1}=S^{z}_1$) for $N=8, 10, 12, 14$ and 16 by the Lanczos mothed.

\begin{figure}
\epsfxsize=40mm 
\centerline{\epsfbox{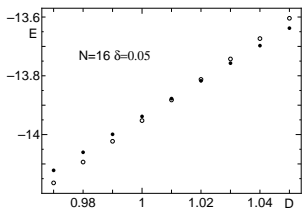}}
\caption{The $D$-dependence of the two lowest energy eigenvalues with twist boundary condition. The energies of the Haldane state and dimer state or large-$D$ state are represented by $\circ$ and $\bullet$, respectively, for $N=16$ and $\delta=0.05$. }
\label{fig1}
\end{figure}

\begin{figure}
\epsfxsize=40mm 
\centerline{\epsfbox{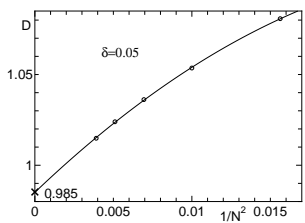}}
\caption{The extrapolation procedure of finite size $\Dc$ for $\delta=0.05$.}
\label{fig2}
\end{figure}

From the results of Tonegawa {\it et. al.}\cite{tt}, it is known that two different kinds of ground states are realized for $D \geq 0$ and for $-1 \leq \delta \leq 1$. For small value of $\delta$ and $D$, the ground state is the Haldane state with a valence bond solid (VBS) structure. Under the twisted boundary condition, the eigenvalues of the space inversion $P$ ($\v{S}_{i}\rightarrow \v{S}_{N-i+1}$) and the spin reversal $T$ ($S_{i}^{z}\rightarrow -S_{i}^{z}, S_{i}^{\pm} \rightarrow -S_{i}^{\mp}$) are all equal to $-1$\cite{ak,kn}. As  $\delta$ and/or $D$ increases, a phase transition takes place to the dimer state or the large-$D$ state for which $P=1$ and $T=1$. We make use of the $P$ and $T$ eigenvalues to distinguish the Haldane phase and dimer or large-$D$ phase. For example, if the value of $\delta$ is fixed, the energies of the two states vary with $D$. For small $D$, the energy of the Haldane state is lower than that of the dimer or large-$D$ state. As $D$ increases, the dimer or large-D state becomes lower than the Haldane state. The two levels cross at one point. This is the finite size transition point $D=\Dc(N)$. Figure \ref{fig1} shows the $D$-dependence of the two lowest levels for $N=16$ and $\delta=0.05$. We extrapolate the critical point as $\Dc(N)=\Dc(\infty)+aN^{-2}+bN^{-4}$. Figure \ref{fig2} shows the extrapolation procedure.

\begin{figure}
\epsfxsize=40mm 
\centerline{\epsfbox{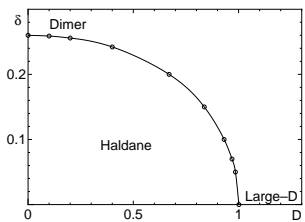}}
\caption{Phase diagram of this model. The Haldane state is inside the solid line, the dimer state and large-$D$ state are outside the solid line.}
\label{fig3}
\end{figure}

In Fig. \ref{fig3}, the ground state phase boundary is represented by the solid line. Inside the solid line, the ground state is the Haldane state. Outside the solid line, the ground state is the dimer or the large-$D$ state. We find  no singularity between the dimer and the large-$D$ state and these two states can be regarded as a single phase. This is in agreement with ref. \citen{tt}
\subsection{Critical behaviour on the phase boundary}
At the critical points, we calculate the energy for the magnetization $M^z=0$ at fixed wave number $k$ by the Lanczos exact diagonalization method under periodic boundary conditions. The magnetization is $M^{z}=\displaystyle\sum_{l=1}^{N}S_{l}^{z}$. The system sizes are $N=8, 10, 12, 14$ and 16. The ground state has $M^{z}=0$ and $k=0$.

\begin{figure}
\epsfxsize=40mm 
\centerline{\epsfbox{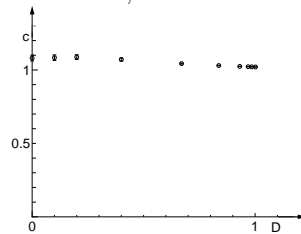}}
\caption{The $D$ dependence of the numerically obtained central charge $c$.}
\label{figc}
\end{figure}

It is known that the finite size correction to the ground state energy is related to the central charge $c$ and the spin wave velocity $v_{\mbox{s}}$ as follows, \cite{ca,HW,IA}
\begin{equation}
\label{eq1}
\frac{1}{N}E_{\mbox{g}}(N) \cong \varepsilon_{\infty}-\frac{\pi cv_{\mbox{s}}}{6N^{2}},
\end{equation}
\begin{equation}
\label{eq2}
v_{\mbox{s}}=\lim_{N \rightarrow \infty}\frac{N}{2\pi}[E_{k_{1}}(N)-E_{\mbox{g}}(N)].
\end{equation}
Here $E_{\mbox{g}}(N)$ is the ground state energy, and $E_{k_{1}}(N)$ is the energy of the excited state with wave number $k_{1}=\frac{2\pi}{N}$ and magnetization $M^{z}=0$. Also, $c$ is the central charge and $\varepsilon_{\infty}$ is the ground state energy per unit cell in the thermodynamic limit. After appropriate extrapolation to $N \rightarrow \infty$, we have checked that the central charge $c$ is close to unity on the phase boundary as shown in Fig. \ref{figc}. Error bars are estimated from the difference between the extrapolated values with the formulae $c(N)=c+C_{1}/N^2$ and $c(N)=c+C_{1}/N^2+C_{2}/N^4$. Therefore, we expect that the present model can be described by a Gaussian model on the critical line,
\begin{equation}
{\cal H_{\mbox{G}}}= \frac{1}{2\pi} \int {{\rm d}x \Big[ v_{\mbox{s}}K (\pi\Pi)^{2}+\frac{v_{\mbox{s}}}{K}(\frac{\partial \phi}{\partial x})^{2} \Big]},
\end{equation}
where $\phi$ is the boson field defined in the interval $0 \leq \phi < \sqrt{2}\pi$, $\Pi$ is the momentum density conjugate to $\phi$ which satisfies $[\phi(x),\Pi(x')]={\rm i}\delta(x-x')$ and $K$ is the Luttinger liquid parameter. Deviation from the critical point is described by the term $\cos \sqrt{2} \phi$. The low energy properties of our model can be described by the one dimensional quantum double sine-Gordon theory near the critical line,
\begin{eqnarray}
\label{SG}
{\cal H_{\mbox{SG}}}&=& \frac{1}{2\pi} \int {{\rm d}x \Big[ v_{\mbox{s}}K (\pi\Pi)^{2}+\frac{v_{\mbox{s}}}{K}(\frac{\partial \phi}{\partial x})^{2} \Big]}\\
&+& \frac{y_{1}v_{\mbox{s}}}{2\pi a^{2}} \int{{\rm d}x \cos \sqrt{2}\phi}+\frac{y_{2}v_{\mbox{s}}}{2\pi a^{2}}\int{{\rm d}x \cos \sqrt{8}\phi}.\nonumber
\end{eqnarray}
The last term becomes marginal at the SU(2) symmetric point ($D=0$). For $D>0$, it becomes irrelevant. We keep this term in the Hamiltonian ${\cal H_{\mbox{SG}}}$ however because it determines the leading finite size correction to the critical dimensions as discussed below.

The scaling dimensions $x_n$ of the operators are related with the energy eigenvalues $E_{n}(N,M^{z})$ of the corresponding excited states as $x_n =\lim_{N \rightarrow \infty} x_n(N)$ where,
\begin{equation}
\label{eq3}
x_{n}(N)=\frac{N}{2\pi v_{\mbox{s}}}[E_{n}(N,M^{z})-E_{\mbox{g}}(N,M^{z})].
\end{equation}
We can identify the correspondence between the operators in boson representation and the eigenstates of spin chains by comparing their symmetry properties.\cite{kn,nomura} Let us denote the scaling dimensions of the operators $\sqrt{2}{\mbox {cos}}\sqrt{2}\phi$ and $\sqrt{2}{\mbox {sin}}\sqrt{2}\phi$ by $x_{1}$ and $x_{2}$, respectively. Both should be equal to $K/2$ in the thermodynamic limit as determined from their correlation functions. Numerically, these exponents are calculated using eq. (\ref{eq3}) for $M^{z}=0$, $P=1$, $k=0 [x_1]$ and for $M^{z}=0$, $P=-1$, $k=0 [x_2]$.  Thus the value of $K$ can be determined from these values. To eliminate the finite size correction due to $y_{2}$, it is more convenient to use the combination,
\begin{equation}
\label{eqkn}
K(N)=x_{1}(N)+x_{2}(N),
\end{equation}
as proposed by Kitazawa and Nomura,\cite{kn} because the leading finite size connections to $x_{1}$ and $x_{2}$ are given by
\begin{equation}
x_{1}=\frac{K}{2}+\frac{y_{2}}{2}(\frac{1}{N})^{2K-2},
\end{equation}
\begin{equation}
x_{2}=\frac{K}{2}-\frac{y_{2}}{2}(\frac{1}{N})^{2K-2}.
\end{equation}

\begin{figure}
\epsfxsize=70mm 
\centerline{\epsfbox{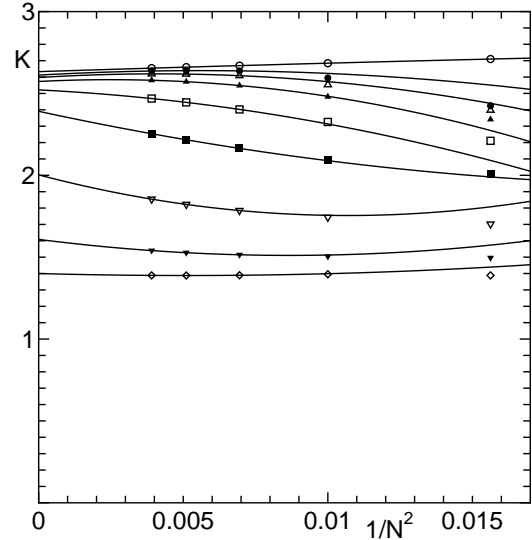}}
\caption{The extrapolation procedure for finite size $K$ with $D=1.001, \delta=0.000; D=0.985, \delta=0.050; D=0.931, \delta=0.100; D=0.837, \delta=0.150; D=0.670, \delta=0.200; D=0.400, \delta=0.242; D=0.200, \delta=0.256$ and $D=0.100, \delta=0.260$ from top to bottom.}
\label{fig4}
\end{figure}

\begin{figure}
\epsfxsize=70mm 
\centerline{\epsfbox{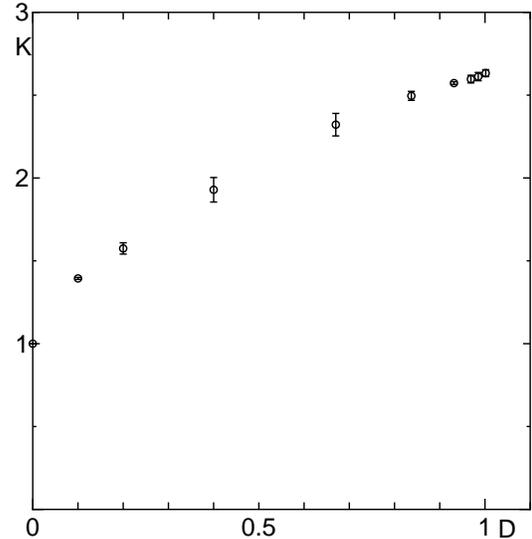}}
\caption{The $D$-dependence of $K$.}
\label{fig5}
\end{figure}

We assume the formula
\begin{equation}
\label{kn}
K(N)=K+\frac{C_{1}}{N^{2}}+\frac{C_{2}}{N^{4}},
\end{equation}
and extrapolate $K(N)$ to $N \rightarrow \infty$. The extrapolation procedure is shown in Fig. \ref{fig4}. The $N$-dependence of $K(N)$ is non-monotonic for small $\delta$. Actually it increases with $N$ for relatively small $N$ but starts to decrease for large $N$. Although such non-monotonic behavior is not observed for large $\delta$, we expect a similar behavior for $N>16$. Therefore we do not include the data for small size systems in the extrapolation procedure of $K(N)$ and only use the values for $N=12,14$ and 16. Further, we regard these extrapolated values as an upper bound while the lower bound is estimated from the value for $N=16$ in the regime $0.15\leq \delta \leq 0.256$. The $D$-dependence of $K$ is shown in the Fig. \ref{fig5}. For $D \simgeq 0.4$, $K$ is larger than 2. This implies that the effective interaction between the elementery fermionic excitation is attractive. In contrast, for $D \simleq 0.4$, $K$ is less than 2 and the effective interaction is repulsive.

Since the second term of eq. (\ref{SG}) produces the gap, the critical exponent $\nu$ of the energy gap is given by
\begin{equation}
\nu=\frac{1}{2-x_1}=\frac{1}{2-\frac{K}{2}},
\end{equation}
\begin{figure}
\epsfxsize=70mm 
\centerline{\epsfbox{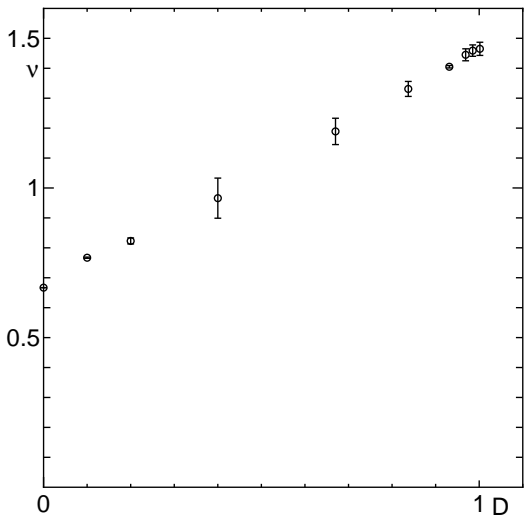}}
\caption{The $D$-dependence of the numerically obtained critical exponent $\nu$.}
\label{fig6}
\end{figure}

The $D$ dependence of the numerically obtained critical exponent $\nu$ is shown in Fig. \ref{fig6}. For $\delta=0, D \cong 1.001$ the value of $\nu$ is $1.465\pm 0.022$. This is consistent with the estimation of Glaus and Schneider\cite{gs} who give $\nu\cong 1.5$, Our estimation procedure is more accurate.

Our model approachs the SU(2) symmetric point in the limit $D \rightarrow 0$. This gives $\nu=2/3$. Our value of $\nu$ is consistent with this and approaches 2/3 in the same limit.

\section{Summary and Discussion}

The ground state phase diagram of a spin-1 antiferromagnetic Heisenberg chain with bond alternation and uniaxial single-ion-type anisotropy is accurately determined by a numerical diagonalization method which uses the twisted boundary condition for $D \geq 0$. The energy gap exponent $\nu$ is calculated by analyzing the numerical diagonalization data using conformal field theory.

\begin{figure}
\epsfxsize=70mm 
\centerline{\epsfbox{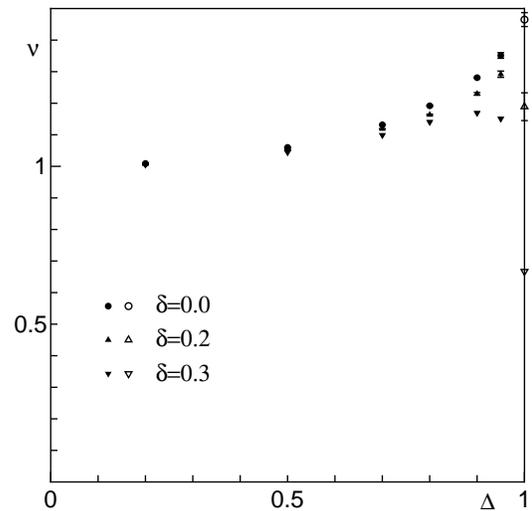}}
\caption{The $\Delta$ dependence of the numerically obtained critical exponent $\nu$ of model (\ref{he}) (filled symbols). The open symbols are the same exponent for the present model (\ref{de}).}
\label{fig8}
\end{figure}

The results of this work are compared with our previous work\cite{wk} on the one dimensional $S=1/2$ XXZ model with dimerization $1-j$ and quadrumerization $\delta$. The Hamiltonian is given by
\begin{eqnarray}
\label{he}
{\cal H} &=& \sum_{l=1}^{2N}  j(S_{2l-1}^{x}S_{2l}^{x}+S_{2l-1}^{y}S_{2l}^{y}+\Delta S_{2l-1}^{z}S_{2l}^{z})   \nonumber \\
& +& \sum_{l=1}^{2N}  (1+(-1)^{l-1} \delta)(S_{2l}^{x}S_{2l+1}^{x}+S_{2l}^{y}S_{2l+1}^{y} \nonumber \\
&+&\Delta S_{2l}^{z}S_{2l+1}^{z}),
\end{eqnarray}
where $\Delta$ is the anisotropy parameter. Figure \ref{fig8} shows the $\Delta$ and $\delta$ dependences of the critical exponent $\nu$. For $\delta< \deltac$, the critical line of this model tends to that of the present model in limit $j \rightarrow -\infty$ and $\Delta \rightarrow 1$ with the product $j(\Delta-1)$ being kept fixed. The product $j(\Delta-1)$ corresponds to $D/2$ in the present model. In Fig. \ref{fig8}, we plot the exponent $\nu$ of the model (\ref{he}) against $\Delta$ for $\delta=0, 0.2$ and 0.3. The open symbols at $\Delta=1$ show the corresponding values of the present model. For $\delta=0$, the exponent $\nu$ monotonically increases with $\Delta$ up to $1.465 \pm 0.022$. For $\delta=0.2$ and 0.3, however, a non monotonic $\Delta$-dependence and strong $\delta$-dependence of $\nu$ near the isotropic point is noted. This is due to the fact that the critical line of the model (\ref{he}) tends to $j \rightarrow -\infty$ in the isotropic limit. For $\delta \geq \deltac$, it decreases down to 2/3.

Our numerical result shows that the value of $K$ varies from 1 to $2.633\pm 0.021$ as $D$ varies from 0 to $1.001 \pm 0.001$. In the neighbourhood of $K=2$ our model can be described by a weakly interacting fermion gas. The explicit construction of an effective weak coupling theory is also a natural extension of our work and is left for future studies.

The numerical calculation is performed using the HITAC S820 and SR2201 at the Information Processing Center of Saitama University and the FACOM VPP500 at the Supercomputer Center of Institute for Solid State Physics, University of Tokyo. This work is supported by a Grant-in-Aid for Scientific Research from the Ministry of Education, Science, Sports and Culture of Japan and a research grant from the Natural Science and Engineering Research Council of Canada (NSERC).

\end{document}


\maketitle

In the numerical calculation to determine the phase boundary, the $D$-term was missing on the boundary site $i=N$. Due to the strong system size dependence caused by this error, the phase boundary and exponents were inaccurate. Although there are corrections to all figures, we only present the corrected  Figs. 1, 2, 3, 6 and 7.  We have checked that  $c$ is closer to unity than the data in original Fig. 4. The data points in Fig. 5 move slightly, but  we do not reproduce it here because Fig. 5 is used only to explain the way of extrapolation. The correction to the open symbols in Fig. 8 is unvisibly small. 
\begin{figure}
\centerline{\includegraphics[width=6cm]{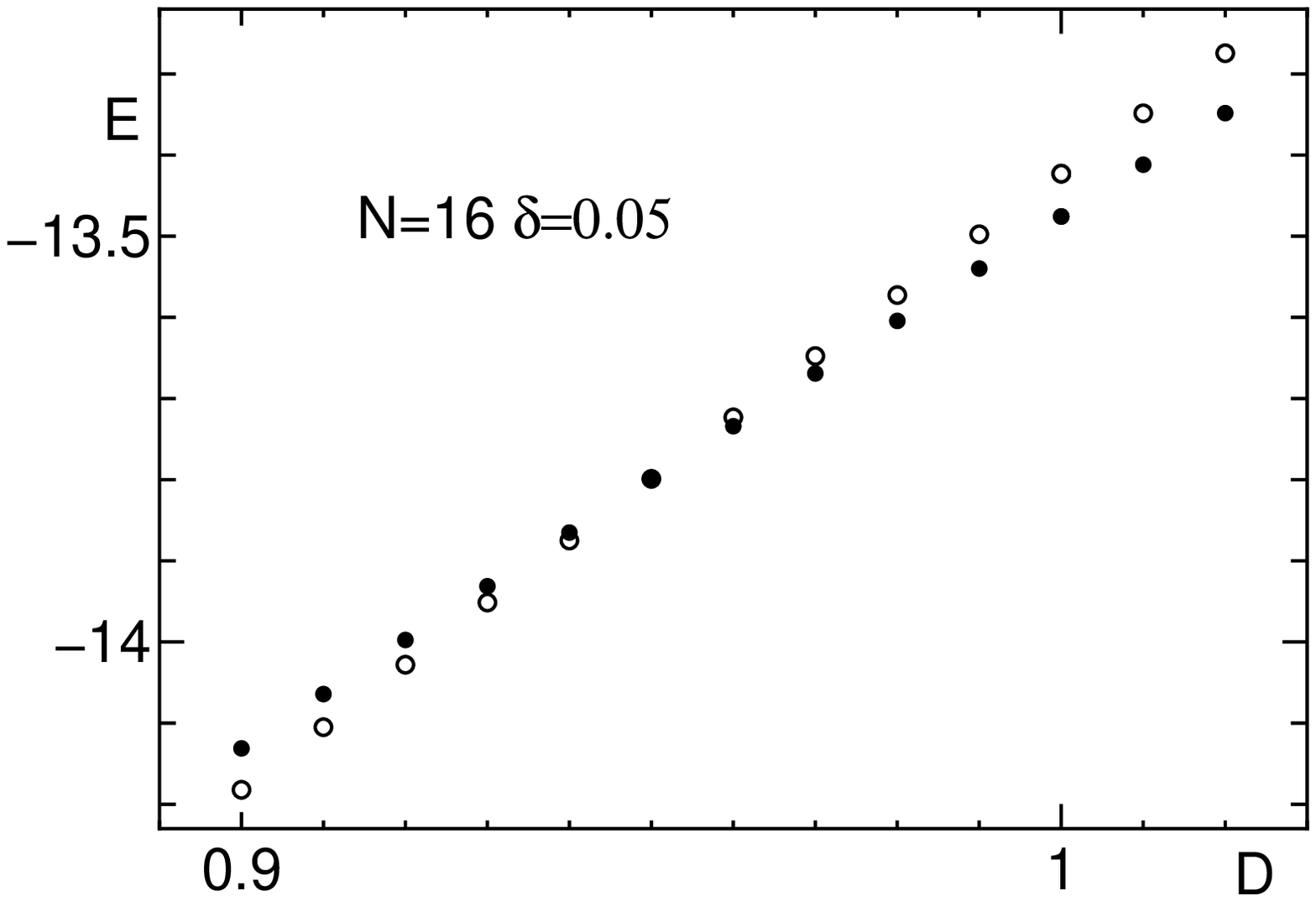}}
\caption{}
\end{figure}
\begin{figure}
\centerline{\includegraphics[width=6cm]{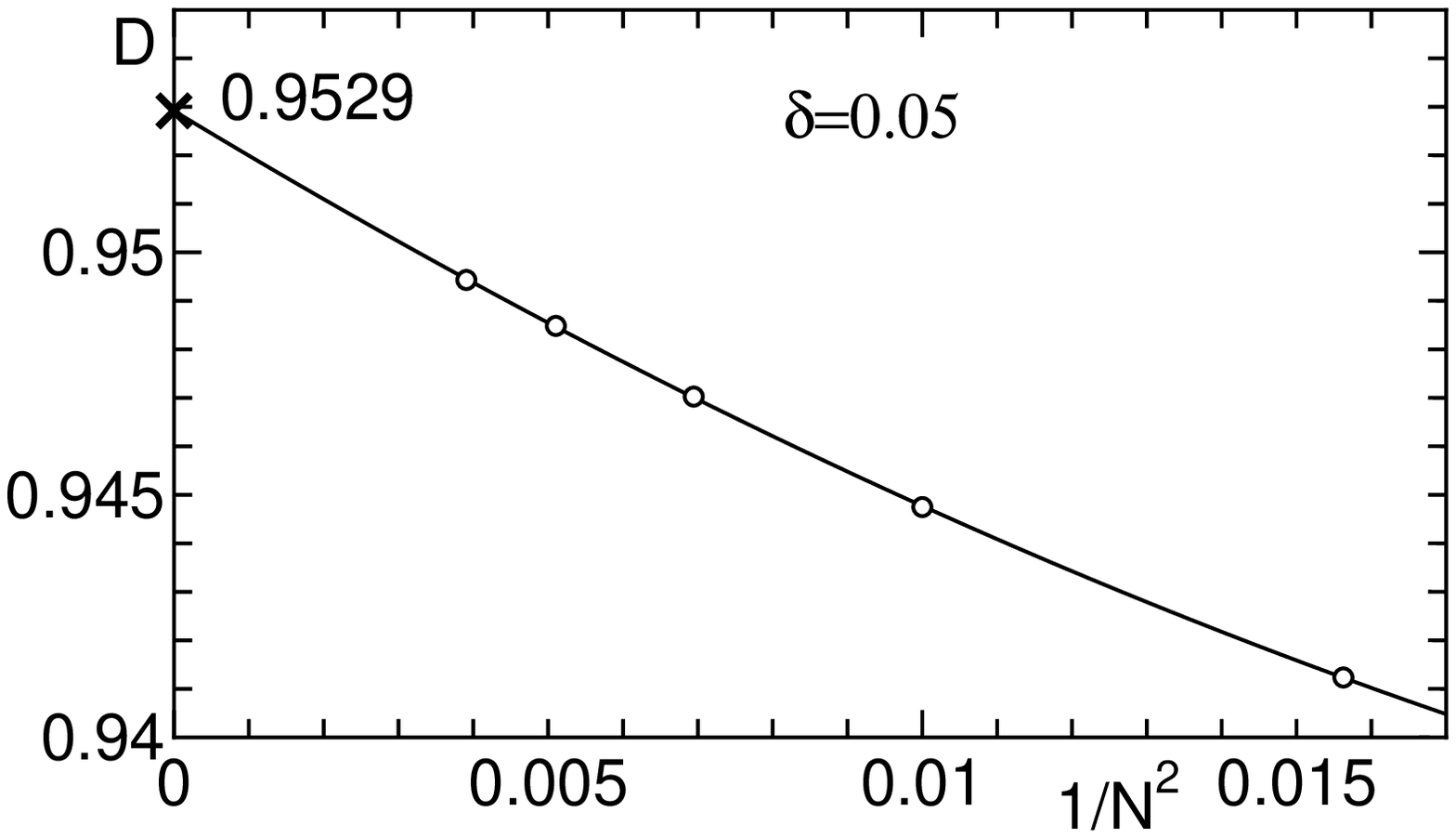}
}
\caption{}
\end{figure}
\begin{figure}
\centerline{\includegraphics[width=6cm]{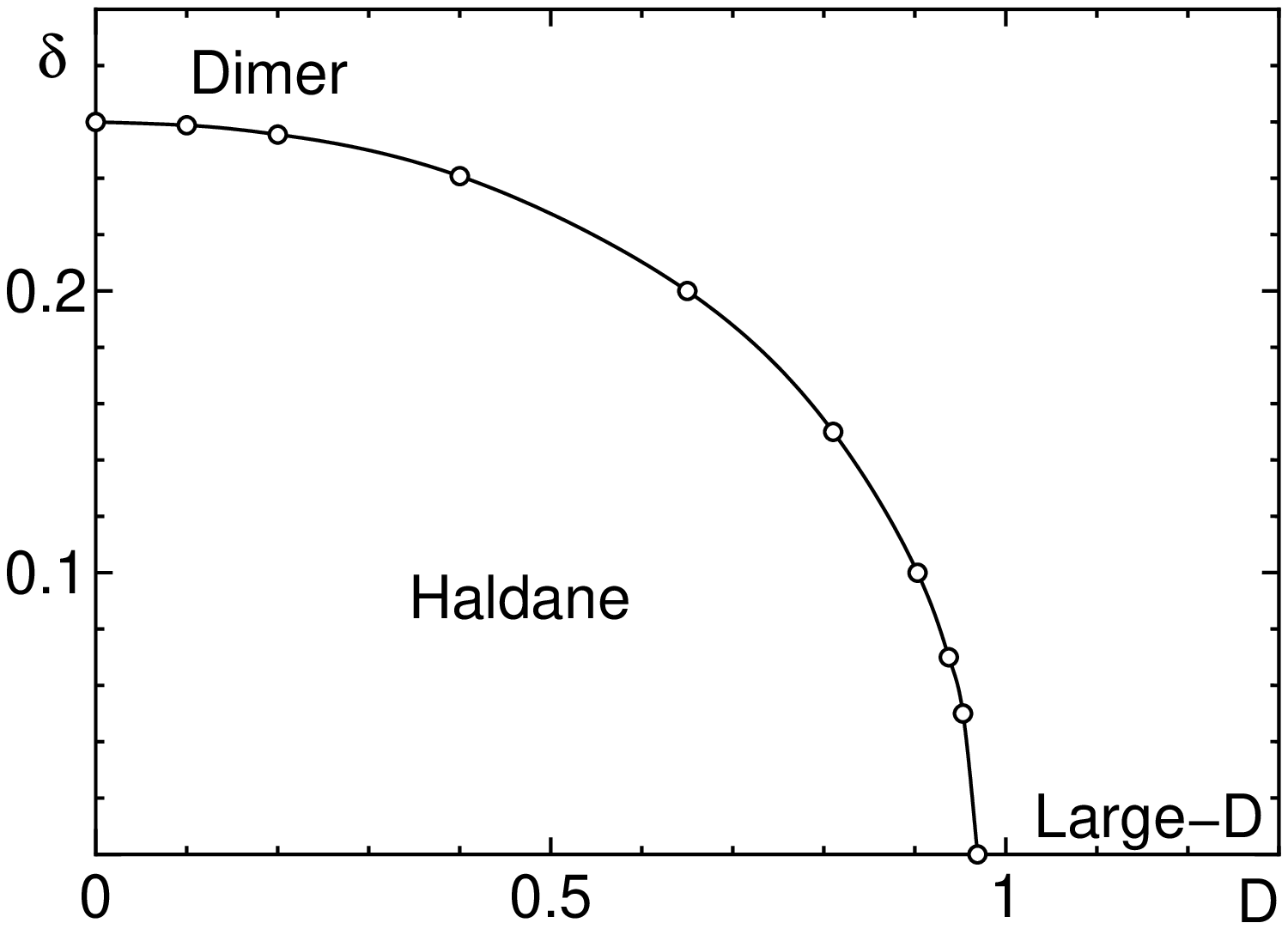}
}
\caption{}
\end{figure}

In the text, the following numerical values should be corrected:  $D_{\rm c}=0.968\pm 0.001$ at $\delta=0$. This value of $D_{\rm c}$  clearly deviates from unity. Therefore the speculation concerning the hidden symmetry at $D=1$ and $\delta=0$ is withdrawn. The exponent  $\nu = 1.4732\pm 0.0004$ at $\delta=0$. The parameter $K$ varies from 1 to $2.6424\pm 0.0003$. The system size $N$ in eqs. (2,2), (2.3) and (2.6) should read $N/2$. 

The authors are grateful to Hiroki Nakano and T\^oru Sakai for  pointing out our error. 

\setcounter{figure}{5}
\begin{figure}
\centerline{\includegraphics[width=6cm]{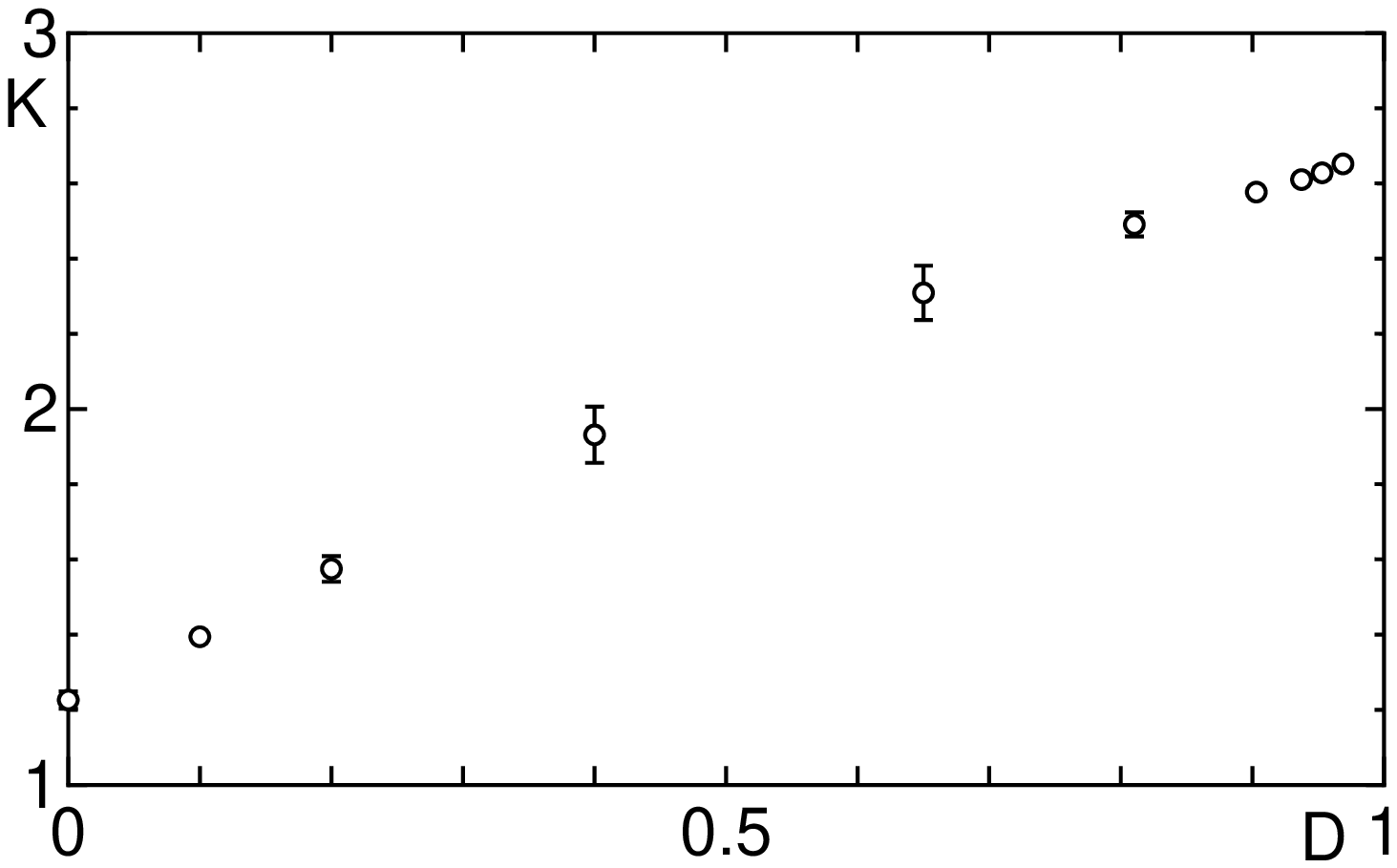}
}
\caption{}
\end{figure}
\begin{figure}
\centerline{\includegraphics[width=6cm]{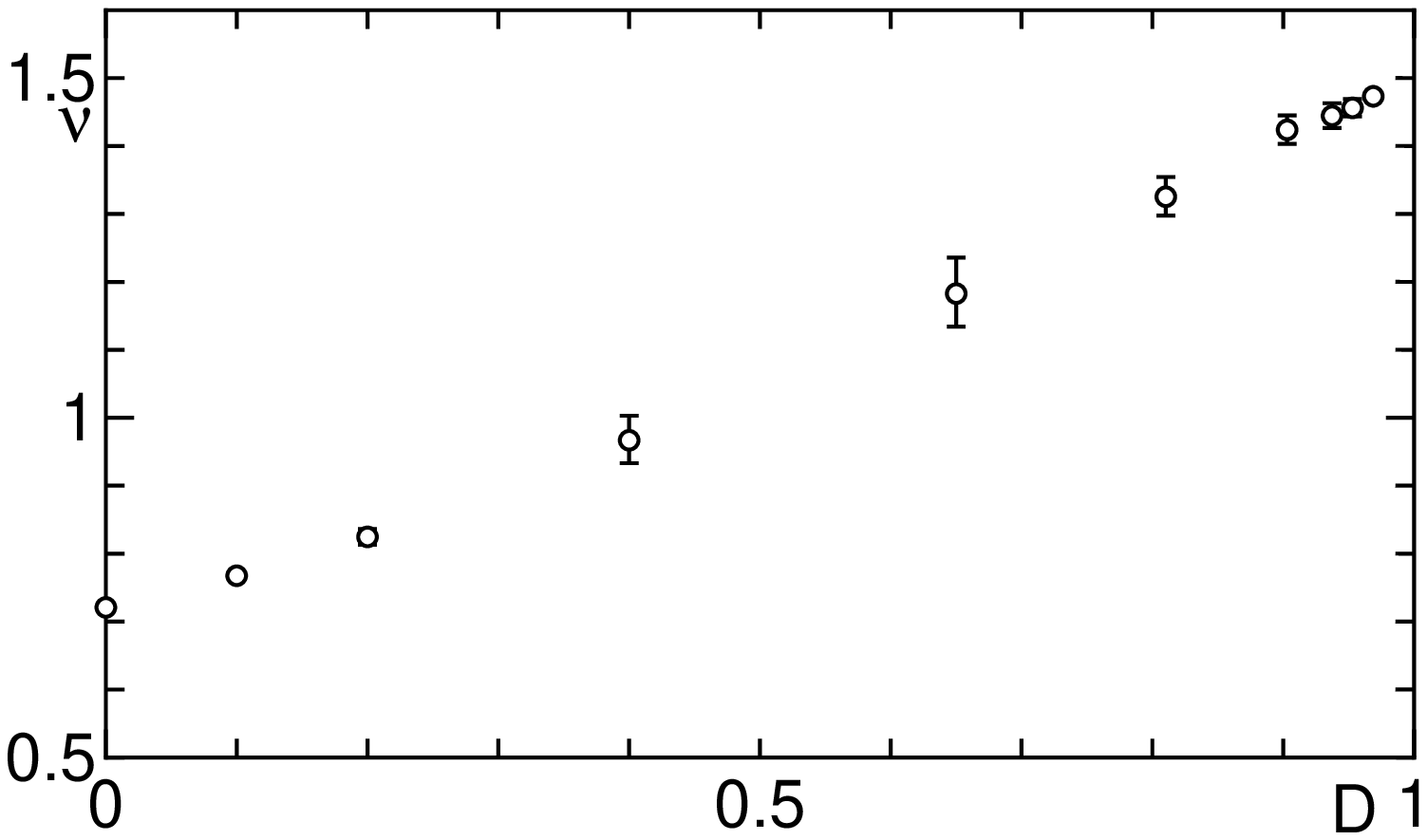}
}
\caption{}
\end{figure}